\documentclass[fleqn,usenatbib]{mnras}

\usepackage{newtxtext,newtxmath}

\usepackage[T1]{fontenc}
\usepackage{ae,aecompl}

\usepackage{graphicx}	
\usepackage{amsmath}	
\usepackage{amssymb}	
\usepackage[dvipsnames]{xcolor}

\title[Cloud model choices and retrievals]{Unveiling cloudy exoplanets: the influence of cloud model choices on retrieval solutions}

\author[Joanna K. Barstow]{
Joanna K. Barstow,$^{1,2}$\thanks{E-mail: jo.barstow@open.ac.uk}
\\
$^{1}$School of Physical Sciences, The Open University, Walton Hall, Milton Keynes, MK7 6AA, UK\\
$^{2}$Department of Physics and Astronomy, University College London, Gower Street, London, WC1E 6BT, UK\\
}

\date{Accepted XXX. Received YYY; in original form ZZZ}

\pubyear{2019}

\begin{document}
\label{firstpage}
\pagerange{\pageref{firstpage}--\pageref{lastpage}}
\maketitle

\begin{abstract}
In recent years, it has become clear that a substantial fraction of transiting exoplanets have some form of aerosol present in their atmospheres. Transit spectroscopy -- mostly of hot Jupiters, but also of some smaller planets -- has provided evidence for this, in the form of steep downward slopes from blue to red in the optical part of the spectrum, and muted gas absorption features throughout. Retrieval studies seeking to constrain the composition of exoplanet atmospheres must therefore account for the presence of aerosols. However, clouds and hazes are complex physical phenomena, and the transit spectra that are currently available allow us to constrain only some of their properties. Therefore, representation of aerosols in retrieval models requires that they are described by only a few parameters, and this has been done in a variety of ways within the literature. Here, I investigate a range of parameterisations for exoplanet aerosol and their effects on retrievals from transmission spectra of hot Jupiters HD 189733b and HD 209458b. I find that results qualitatively agree for the cloud/haze itself regardless of the parameterisation used, and indeed using multiple approaches provides a more holistic picture; the retrieved abundance of H$_2$O is also very robust to assumptions about aerosols. I also find strong evidence that aerosol on HD 209458b covers less than half of the terminator region, whilst the picture is less clear for HD 189733b.

\end{abstract}

\begin{keywords}
radiative transfer -- planets and satellites: atmospheres -- techniques: spectroscopic
\end{keywords}



\section{Introduction}
In recent years, the characterisation of transiting exoplanet atmospheres has evolved to the extent that several comparative studies of the most favourable targets have been published. These targets mostly consist of hot Jupiters, which are ideal candidates for transit spectroscopy. Their high temperatures and H$_2$-He dominated atmospheres result in large atmospheric scale heights, and therefore large fluctuations in the transit depth as a function of wavelength. 

Evidence for the presence of cloud or haze (or a lack of evidence for its absence) has been found in the majority of hot Jupiter spectra. Whilst simulations including condensational processes or cloud microphysics can predict the cloud that is expected to form under particular circumstances, we have no prior knowledge of likely cloud structures on intensely irradiated, hot worlds. Therefore, simple, parameterised retrieval models (e.g. \citealt{lee12, line13a, madhu09, waldmann15, benneke15, cubillos17, blecic17, kitzmann19, mai19,molliere20}) are a crucial part of a data-driven approach to learning about exoplanet aerosols. 

Cloud refers to aerosol produced via condensation, whereas haze is aerosol that is photochemically produced (see e.g. \citealt{horst17}). We do not yet know which mechanism is responsible for aerosol production on exoplanets, but in general discussion about likely constituents centres around condensates (e.g. \citealt{wakeford17} and \citealt{helling16}; the latter elaborates on the full complexity of the process and the need for seed particles). Given the emaphasis on condensation, in this work I will refer to exoplanet aerosols as cloud, as is commonly done in the literature.

Four comparative retrieval studies have been recently published, two dealing with data from the \textit{Hubble Space Telescope} Wide Field Camera 3 (WFC3) instrument only \citep{tsiaras18,fisher18}, and two also incorporating \textit{Hubble} Space Telescope Imaging Spectrograph (STIS) and \textit{Spitzer} InfraRed Array Camera (IRAC) data \citep{barstow17,pinhas19}. The general philosophy of these studies is the same; to use a relatively agnostic model set up to retrieve the basic atmospheric properties for a range of planets, in such a way that the results are directly comparable. Retrieved properties common to all studies include H$_2$O abundance; temperature structure (with varying assumptions); radius or pressure baseline; and cloud parameters. A detailed comparison of the retrieval results from these studies is included in \cite{barstow20}. Of particular interest in this work are the differences in cloud parameterisation, and the effects of this on the retrieved cloud properties. 

In this work, I contrast the different approaches to parameterising cloud, and test each of the cloud models in turn on a benchmark dataset. For this purpose, I have chosen the spectra of HD 189733b \cite{bouchy05} and HD 209458b \cite{charbonneau00} as presented in \cite{sing16}. These are the most precisely measured hot Jupiter spectra so far, and both contain clear evidence for scattering particles in the atmosphere of the planet, albeit of different kinds. Whilst the HD 189733b spectrum has a substantial slope in the visible, which may be an indicator of the presence of small, Rayleigh scattering particles, the HD 209458b spectrum has a muted H$_2$O feature but very little optical slope, suggesting its spectrum is more likely to be dominated by larger particles. Together, these spectra provide a good test of the ability of each cloud model to match a diverse range of observations. 

Previous work by \cite{mai19} applies a similar approach, for more physically-motivated cloud parameterisations, to simulated \textit{James Webb Space Telescope} spectra of WASP-62b. They conclude that including cloud in the model is necessary for retrieving spectra of a cloudy planet, but that all cloud models they tested provide unbiased solutions for other atmospheric properties, despite the differences in approach.

In Sections~\ref{nemesis} and~\ref{basic} I describe the basic retrieval model setup. In Section~\ref{cloud_models}, I discuss the differences between the cloud models used in each paper, and the way in which they have been implemented. Retrieval results are presented in Section~\ref{results}, and discussed further in Section~\ref{discussion}.  

\section{Retrieval model set up}
\subsection{NEMESIS}
\label{nemesis}
I use the NEMESIS radiative transfer and retrieval code, originally developed for Solar System planets \citep{irwin08} and subsequently extended for use with exoplanets \citep{lee12,barstow17,krissansen-totton18}. NEMESIS incorporates a fast, 1D radiative transfer calculation, which makes use of the correlated-k approximation allowing absorption line data to be stored in a quick look up table \citep{lacis91}. Whilst originally NEMESIS used an Optimal Estimation technique to converge on the most likely atmospheric solution \citep{rodg00}, the dependence of this method on an informative prior means that it is less appropriate for exoplanets since in this case our prior knowledge is severely restricted. Instead, a recent upgrade \citep{krissansen-totton18} sees NEMESIS interface with the PyMultiNest algorithm \citep{feroz08,feroz09,feroz13,buchner14}. 

\subsection{Basic atmospheric retrieval model}
\label{basic}
The basic atmospheric models for HD 189733b and HD 209458b are based on those used in \cite{barstow17}. We restrict the spectrally active gases to Na, K and H$_2$O, as these are the only gases for which strong evidence has been found, with prior ranges for $\log$(Na,K) of -13 to -3, and for $\log$(H$_2$O) of -8 to -2. They are assumed to be well mixed, so gas abundances account for three parameters in the retrieval. The bulk of the atmosphere is a mixture of H$_2$ and He. 

Line data for H$_2$O are taken from the BT2 database \citep{barber06}, and those for Na and K are from the VALD database \citep{heiter08}. Collision-induced absorption for H$_2$ and He is taken from \cite{borysow89,borysowfm89,borysow90,Borysow2001,borysow02}. The Na and K features are expected to be strongly pressure-broadened in the wings of the absorption lines. To account for this, the line wing cutoff for Na and K is extended to 6000 cm$^{-1}$ from the line centre, as opposed to the more usual 25 cm$^{-1}$. \cite{welbanks19b} explore the H$_2$-broadened shape of these lines further, and find that at temperatures of 2000 K the wings of both Na and K features do not extend beyond 1.4$\upmu$m; cutting at 6000 cm$^-1$ for the K band means the line extends to 1.41 $\upmu$m. In any case, for cloudy planets the pressure-broadened line wings are generally obscured by the cloud, which can be seen to be the case here.

The temperature profile is represented as an isotherm for $P < 0.1$ atm and $P > 1.0$ atm, and as an adiabat in between. The stratospheric temperature $T_{\mathrm{strat}}$ (prior range: 100---3000 K) is therefore the single temperature variable in the model. I also retrieve the planetary radius at the 10 atm pressure level, because the literature value for this is taken from the white light transit, and the pressure that this refers to is extremely dependent on the atmospheric properties of the planet. For further context as to why this is necessary, see \cite{barstow20} Section 1.2 and references therein. The prior range for this value is planet specific, from 0.9---1.3 R$_{\mathrm{J}}$ for HD 189337b, and from 1.1---1.5 R$_{\mathrm{J}}$ for HD 209458b.  Therefore, before cloud is introduced, the model atmosphere contains 5 free parameters. 

\subsection{Cloud models}
\label{cloud_models}
Cloud formation is a complex process, and reducing clouds to something that can be represented by a minimal number of parameters is challenging. As a result, different teams have adopted a variety of approaches, but all have some factors in common. 

The key features of a cloud that are important to capture when modelling transit spectra are 1) the pressure level at which the cloud becomes opaque and 2) the wavelength dependence of this. These effects can be represented in a variety of ways. Additional effects may also be important under certain circumstances; if a cloud layer is optically thin, the location of the cloud base may also become important. For hot Jupiters, which are tidally locked, strong winds are thought to result in the temperature varying considerably between the morning and evening terminators, which in turn could lead to variation in cloud coverage (e.g. \citealt{line16}); it may therefore also be necessary to include some formulation for fractional cloud cover around the terminator. The approaches to representing these key cloud features vary between the four comparative studies, and are summarised below. 

\cite{barstow17} (hereafter B17) use a grid of cloud models. They assume that the cloud has uniform specific density for pressures above a variable cloud top pressure. The cloud is treated either as Rayleigh scattering (scattering efficiency scales as 1/$\lambda^4$, representing small particles) or grey (constant scattering efficiency). Two vertical cloud distributions are tested - the cloud either extends from the top pressure to the bottom of the atmosphere, or extends downwards by a decade in pressure. The second approach allows a detached haze layer to be simulated as well as a deep cloud deck. Finally, total optical depth is a free parameter in an Optimal Estimation retrieval for each cloud model.

In this work, the B17 model has been extended from the grid-search approach to include four free parameters: total nadir optical depth $\tau$; top pressure $P_{top}$; base pressure $P_{base}$; and scattering index $\gamma$, where the wavelength dependence of the extinction efficiency is proportional to $\lambda^{-\gamma}$.

\cite{tsiaras18} (T18) and \cite{fisher18} (F18) use effectively the same cloud model as each other, based on that presented by \cite{kitzmann18}, which in turn evolved from models used by \cite{lee13} and \cite{lavie17}. They use an analytical model to capture the functional dependence of cloud extinction on wavelength. As described in \cite{fisher18}, the cloud opacity is parameterized as follows:

\begin{equation}
    \kappa_{\mathrm{cloud}} = \frac{\kappa_0}{Q_0 x^{-a} + x^{0.2}}
\end{equation}

where $\kappa_0$ is an optical depth scaling factor, $Q_0$ determines the wavelength at which the extinction efficiency peaks, $a$ is a scattering slope index and $x$ is the particle size parameter, given by

\begin{equation}
    x = \frac{2{\pi}r}{\lambda}
\end{equation}

where $r$ is the effective particle radius and $\lambda$ is the wavelength. The model has four free parameters, $\kappa_0$, $Q_0$, $a$ and $r$. The $Q_0$ parameter is of particular interest because the wavelength at which the extinction efficiency peaks is related to the composition of the aerosol particles, so retrieval of this parameter could provide some constraint on possible cloud species.

The model presented by \cite{kitzmann18} could be extended to include constraints on the vertical location of the cloud; however, here I adopt the version of the model used in F18, which assumes the cloud is spread vertically throughout the atmosphere.

T18 use a special case of this formalism, which fixes $Q_0$ to 50 and $a$ to 4. They fix $\kappa_0$ to 5, but additionally retrieve a cloud mixing ratio $\chi_C$ which scales the total opacity in the same way. This model is combined with a simple opaque, grey cloud for pressures $P$ > $P_{\mathrm{top}}$. 

Finally, \cite{pinhas19} (P19) include a cloud model combining an optically thick grey cloud (as used in T18 for $P$ > $P_{\mathrm{top}}$) with an overlying haze layer, parameterized by an optical depth scaling and a power law dependence of extinction with wavelength. This follows on from the model introduced in \cite{macdonald17}:

\begin{equation*}
\textit{P}< P_{top}:~~~\kappa = n_{H_2} a \sigma_0(\lambda/\lambda_0)^{-\gamma}
\end{equation*}

\begin{equation*}
\textit{P}> P_{top}:~~~\kappa = \infty
\end{equation*}

where $n_{H_2}$ is the number density of H$_2$ molecules, $\sigma_0$ is the molecular scattering opacity at reference wavelength $\lambda_0$, and $a$ here represents the `Rayleigh enhancement factor' describing the magnitude of additional scattering due to clouds. $\gamma$ is the dependence of the scattering efficiency on wavelength $\lambda$.

\begin{figure*}
	\centering
	\includegraphics[width=\textwidth]{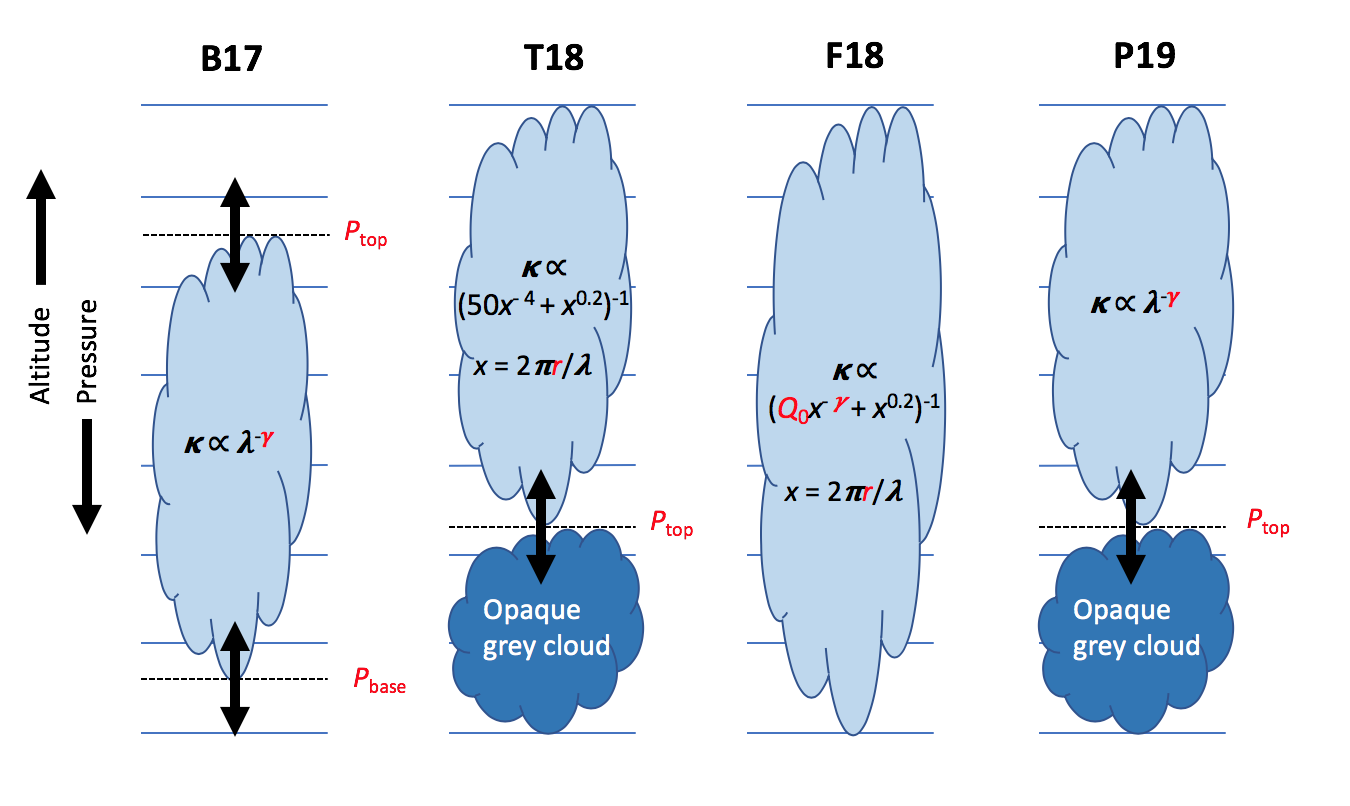}
    \caption{This provides a visual indication of how cloud structure is parameterised in each of the models discussed. The parameters highlighted in red are those that are allowed to vary in the retrieval.}
    \label{cloudparams_fig}
\end{figure*}

All of these four possible parameterizations have been incorporated into the updated version of the \textit{NEMESIS} spectral retrieval code that works with PyMultiNest.  The free and fixed parameters in each of the models considered, as set in their respective papers, are summarized in Table~\ref{cloudparams} and Figure~\ref{cloudparams_fig}. These were incorporated into \textit{NEMESIS} as far as possible in the same way as in the original papers, but the results have been processed such that they can be compared across models. For example, instead of presenting opacity scaling factors (which differ in meaning between the four models), after the retrieval I combine the relevant parameters from each model to calculate the nadir optical depth at 0.2 $\upmu$m, and present this value. For the B17 and P19 models, the optical depth is directly equivalent to the scaling factor due to the fact that the scattering is a simple power law and the cross section is normalised to 1 at 0.2 $\upmu$m; for T18 and F18, the scattering cross section depends on particle size and is not normalised.

The prior ranges for each case, as incorporated into NEMESIS, are also included in Table~\ref{cloudparams}. In general, very wide priors have been used to limit retrieval dependence on the prior, with a few exceptions. The opacity scaling factor range is slightly narrower for T18 and F18 because this does not directly correspond to an optical depth, and keeping the upper end of the range the same as for B17 and P19 resulted in eventual optical depths that were larger than NEMESIS could cope with; however, these ranges are still extremely wide. The other exceptions are for the F18 model, which uses prior ranges for the scattering index and shape factor suggested by the discussion of the model in \cite{kitzmann18}. This model is derived from an analytical fit to Mie scattering calculations, and the authors found that typical scattering indices were between 3 and 7, and the shape factor for the various possible species varied from 0.07 to 64.98.

I also test options for all cloud models to incorporate a fractional cloud coverage parameter. This assumes that some fraction $f$ of the terminator has cloud, while the remainder (1-$f$) is entirely cloud-free. P19 and \cite{macdonald17} already include a fractional cloud parameter in their analysis; here, I investigate whether the inclusion of an extra free parameter is justified by the information content of current data. 

\begin{table*}
	\centering
	\caption{Variable cloud parameters grouped by type, for each of the models considered. Quantities are as defined in the text. Prior ranges as used in the retrieval are also included.}
	\label{cloudparams}
	\begin{tabular}{lcccc} 
		\hline
		Property & B17 (updated) & T18 & F18 & P19\\
		\hline
		Opacity & $\tau$ & $\chi_c$ & $\kappa_0$ & $a$ \\
		 & 10$^{-10}$, 10$^{20}$ & 10$^{-10}$, 10$^{13}$ & 10$^{-10}$, 10$^{13}$ & 10$^{-10}$, 10$^{20}$\\
		Scat index & -$\gamma$ & -4 & -$a$ & -$\gamma$ \\
		 & 0, 14 & Fixed & 3, 7 & 0, 14\\
		Top pressure & $P_{top,all}$ & $P_{top,grey}$ & None & $P_{top,grey}$\\
		 & 10$^{-8}$, 1.0 & 10$^{-8}$, 1.0 & N/A & 10$^{-8}$, 1.0\\ 
		Base pressure & $P_{base}$ & None & None & None\\
		 & $P_{top}$, 1.0 & N/A & N/A & N/A\\
		Particle size & None & $r$ & $r$ & None\\
		 & N/A & 10$^{-3}$, 10$^{2}$ & 10$^{-3}$, 10$^{2}$ & N/A\\
		Shape factor & None & 50& $Q_0$ & None \\
		 & N/A & Fixed & 0.1, 65 & N/A\\
		\hline
	\end{tabular}
\end{table*}

\section{Retrieval results}
\label{results}
Here I present and compare the results for each model, for HD 189733b in Section~\ref{hd189_results} and for HD 209458b in Section~\ref{hd209_results}. With one exception, all four cloud parameterizations can produce a fit to the data for all scenarios, and the results are generally consistent. I find that there is good evidence in favour of including a terminator cloud fraction parameter for HD 209458b, but a more confused picture for HD 189733b. 

A selection of retrievals are shown within the paper, and the full results can be found in an online repository\footnote{https://tinyurl.com/qvpazxw}. All corner plots are generated using the corner.py routine \citep{corner}. 

Both B17 and P19 found that the H$_2$O abundances for these hot Jupiters were generally below that expected for a solar composition gas under the conditions of HD 189733b and HD 209458b. The disequilibrium chemistry, solar composition models of \cite{moses11} predict log(H$_2$O) values of -3.42 and -3.45 for HD 189733 and HD 209458b respectively. I compare these with the values retrieved in this study below. 

\subsection{HD 189733b}
\label{hd189_results}
The spectrum of HD 189733b displays a very steep slope at visible wavelengths, and a muted but still present H$_2$O feature in the near infrared. The steep visible slope has generally been attributed to the presence of high altitude clouds, and the results presented here corroborate this assumption. 

The extreme steepness of the visible slope presents a challenge to a completely agnostic approach to retrieval, which aims to place minimal prior constraints on the solution. A very steep slope of this kind can be produced in two ways; either by modelling the cloud such that the extinction efficiency drops off very rapidly with wavelength, or by increasing the temperature and thence the scale height of the atmosphere, which increases the amplitude of all features. 

For HD 189733b, if the temperature is allowed to vary freely up to a threshold of 3000 K, then the retrieved terminator temperature is around 2000 K for all models. However, we actually have some information about the range of values the terminator temperature can take, since we know the maximum equilibrium temperature of HD 189733b (assuming zero albedo) is around 1200 K \citep{sing16,barstow17}. In transit, for low spectral resolution, the region of the atmosphere that is probed at visible to IR wavelengths typically covers the range 10$^{-4}$ --- 0.1 bar \citep{welbanks19}, which we expect to be broadly isothermal. Therefore, the temperature in the regions of the terminator to which we are sensitive should be $\le$ 1200 K, so we can place a (conservative) informative prior on the temperature of 100---1300 K. 

\begin{figure*}
	\centering
	\includegraphics[width=\textwidth]{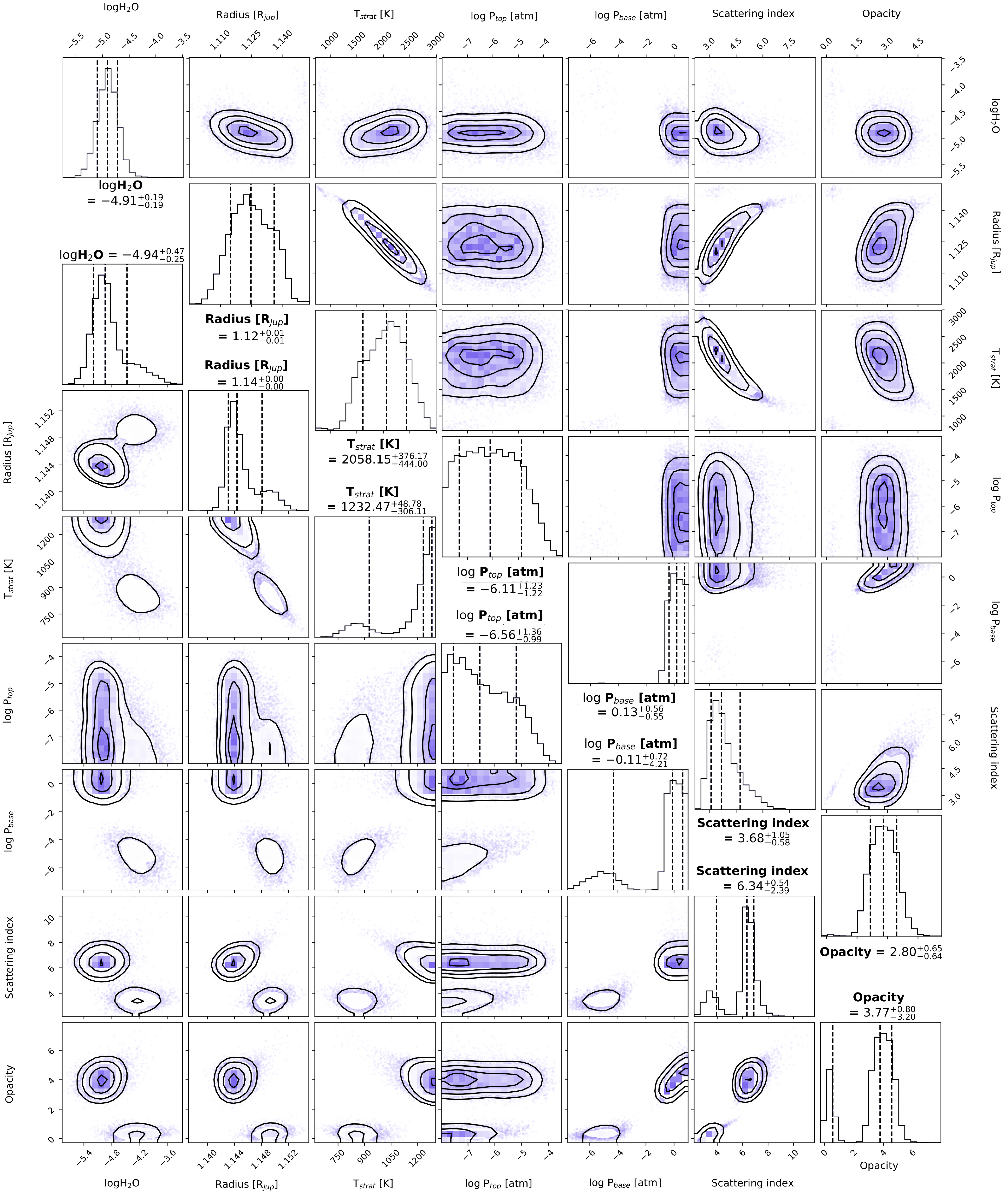}
    \caption{This figure compares posterior corner plots for the B17 cloud model with (bottom left) and without (top right) the informative temperature prior. The secondary solution in the bottom left plot corresponds to the solution presented in B17, arrived at using a more restricted version of the model.}
    \label{hd189_b}
\end{figure*}

\begin{figure*}
	\centering
	\includegraphics[width=\textwidth]{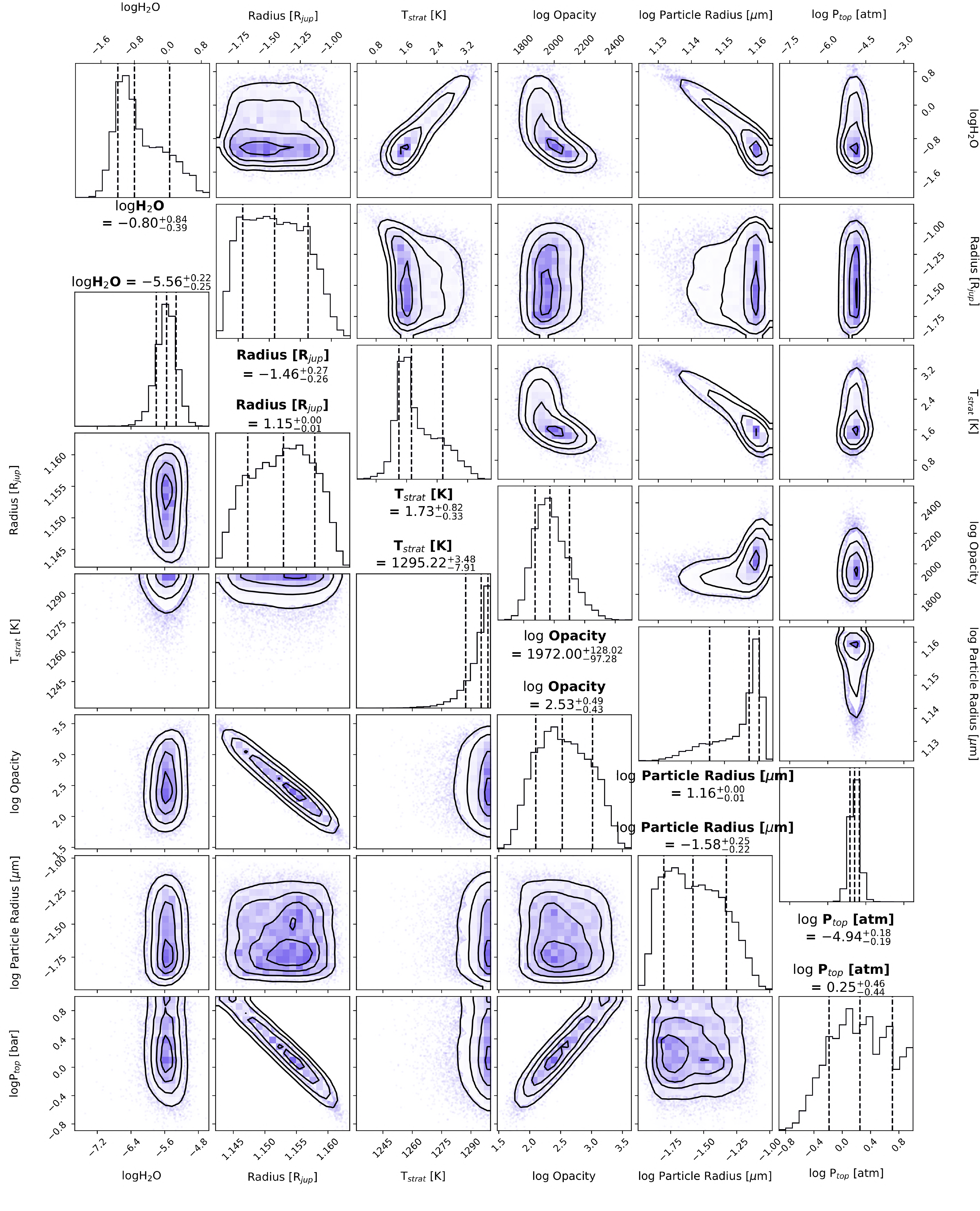}
    \caption{As Figure~\ref{hd189_b} but for the T18 model}
    \label{hd189_t}
\end{figure*}

\begin{figure}
	\centering
	\includegraphics[width=\columnwidth]{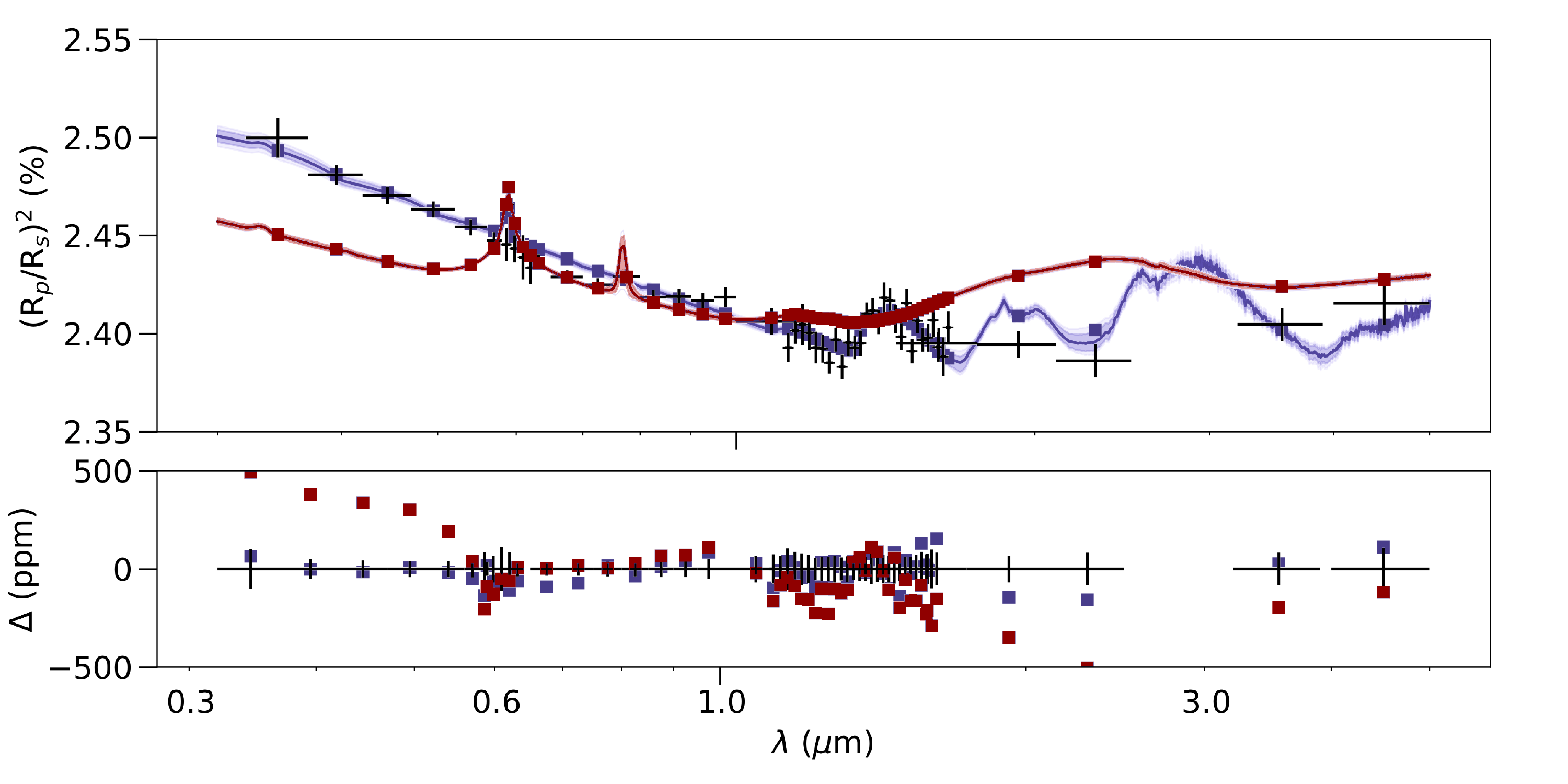}
    \caption{Spectral fits to the HD 189733b data (red) with the T18 model with a restricted temperature prior (dark red) and a broad prior (purple). Spectra are generated using the median values of the posterior distributions. The fit is generally poor for the restricted prior case as the relatively inflexible cloud parameterisation does not allow the short wavelength  slope to be matched. Shaded regions on the spectrum plot indicate the 2-$\sigma$ envelope for each case}
    \label{tsiaras_fits}
\end{figure}

I compare the results of the scenarios with and without the informative temperature prior for the B17 and T18 models (Figures~\ref{hd189_b} and ~\ref{hd189_t}). In the case where the retrieved temperature is allowed to exceed the 1300 K limit, the steep visible slope is fit by increasing the scale height via an increase in temperature; for cases where the temperature is restricted, this is compensated for by the scattering index parameter in B17, with the scattering index increasing between the two cases by around 2.75. The T18 model does not allow for a tuneable scattering index, and the result of this is that for the restricted temperature case the T18 model is unable to produce an adequate fit to the data (Figure~\ref{tsiaras_fits}). This is not a criticism of the T18 study itself, as it dealt only with data from \textit{Hubble}/WFC3 and the model was not therefore required to simulate a steep optical scattering slope. However, this demonstrates that fixing the values of model parameters should be undertaken with caution. 

In the restricted temperature case, temperatures at the high end of the prior range are still favoured. This begs the question as to why the retrieval converges preferentially on solutions involving high temperatures rather than cases with a steep scattering slope produced by the cloud. I suggest that the reason for this is that the high temperature solution for the large optical slope relies only on a single model parameter, whereas the solutions involving cloud (regardless of the cloud model chosen) rely on a specific combination of at least three parameters. Occam's razor would therefore favour the temperature solution, and this is only rejected in the light of the prior knowledge we have of the planet's temperature.

I adopt the restricted temperature case as the more plausible scenario, and key retrieved results are shown in Table~\ref{table_189_results}. 

Other points of interest from these results are the consistency of the H$_2$O abundance results, and the retrieved scattering index, across B17, F18 and P19 for the homogeneous cloud case. All models indicate an H$_2$O abundance of approximately 1/30$\times$ solar, consistent with previous studies B17 and P19, and cloud with a scattering index of around 6.4. Rayleigh scattering corresponds to an index of 4, so the cloud particles present display super-Rayleigh behaviour and are likely to be small. This is discussed further in Section~\ref{scattering}. 

The retrieved cloud top pressures must be compared with more caution, as this parameter represents different things across the four models. In B17 it represents the top of the entire cloud deck, whereas in T18 and P19 it is the top of the grey, opaque cloud. Therefore, the retrieval of a very low top pressure in B17 compared with the others is consistent. 

In summary, all models for HD 189733b indicate the presence of small particle cloud high in the atmosphere, and absence of an opaque grey cloud, and sub-solar H$_2$O. 

After performing the initial retrievals including homogeneous cloud, I extend the analysis to also include a cloud fraction parameter for all cloud models. Key results from this analysis are presented in Table~\ref{table_189_results}, along with the log of the Bayesian evidence for each model relative to the model with the highest evidence. This is also known as the Bayes factor. Generally, if the Bayes factor difference is $>$ 2 the model with the higher evidence is moderately favoured; if $>$ 5 then the higher evidence model is strongly favoured.  It can be seen that the model with the highest evidence for HD 189733b is the P19 model including heterogeneous cloud. The improvement factor for the P19 model when the cloud fraction parameter is added is 6.3, indicating that the heterogeneous cloud is strongly favoured for the P19 model; however, similar improvements when fractional cloud is included are not seen for the other parameterisations, with weak to no evidence for including fractional cloud.

Comparing the different parameterisations against each other, we see that for the fractional cloud case the P19 model significantly outperforms all others. However, if fractional cloud is not included, the B17, F18 and P19 models perform similarly.

\begin{figure}
	\centering
	\includegraphics[width=1.1\columnwidth]{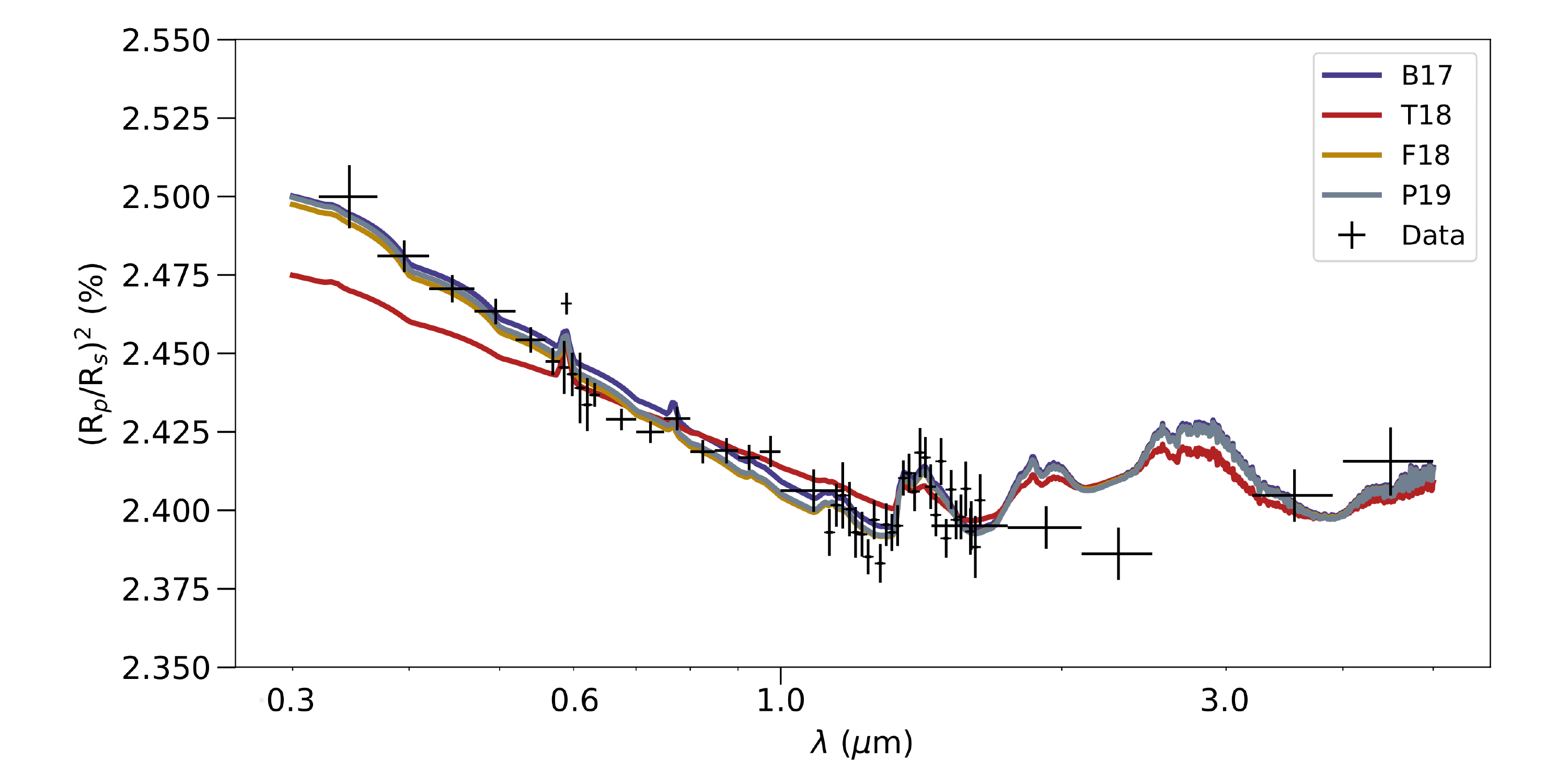}
    \caption{Spectral fits to the HD 189733b data for all models, for the restricted temperature case without fractional cloud. The model spectra are calculated using the median parameter values.}
    \label{hd189_fits}
\end{figure}

\begin{figure}
	\centering
	\includegraphics[width=1.1\columnwidth]{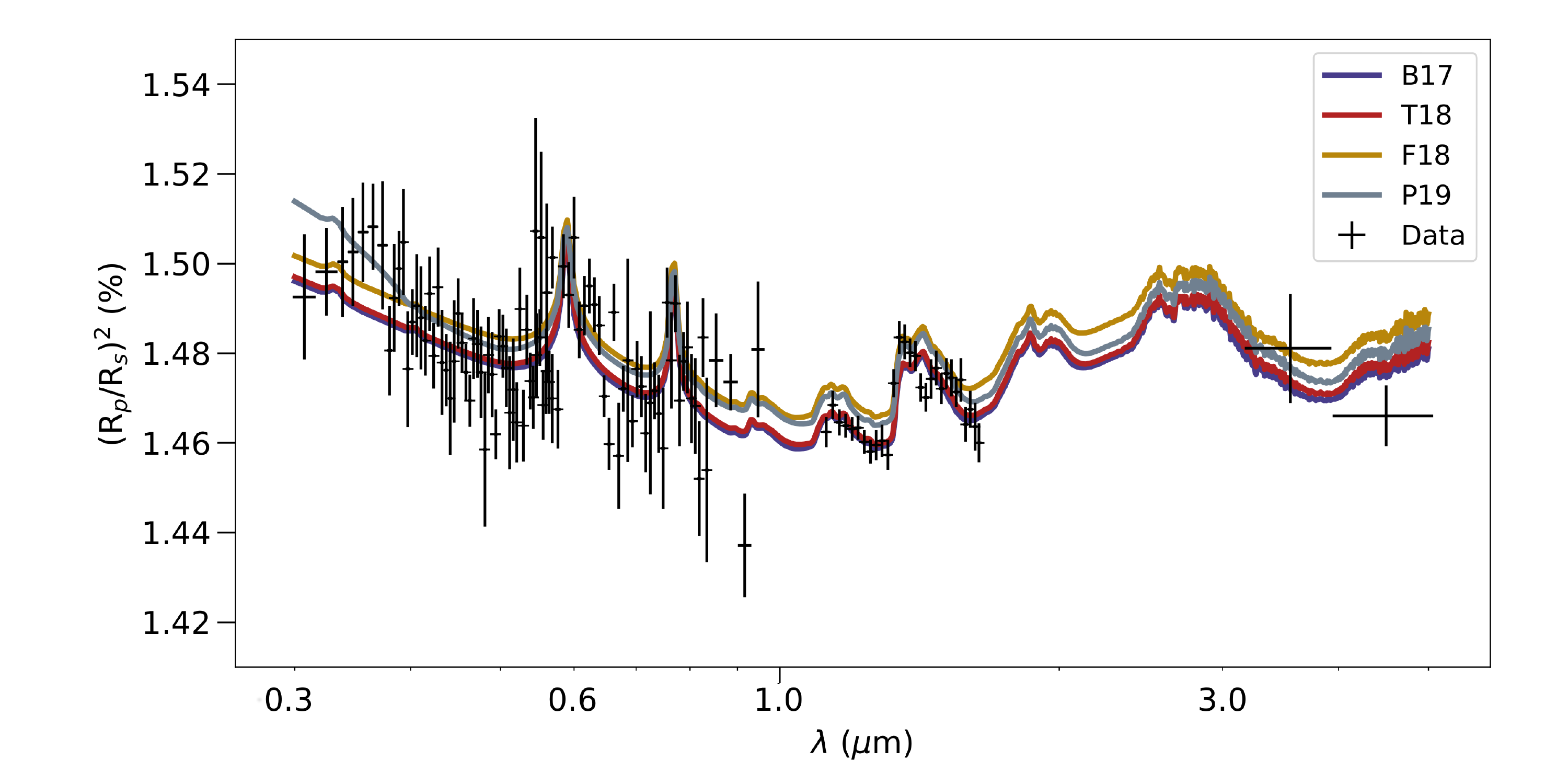}
    \caption{As Figure~\ref{hd189_fits} for HD 209458b.}
    \label{hd209_fits}
\end{figure}

\begin{table}
	\centering
	\caption{Key retrieval results for HD 189733b. The difference in ln(Bayesian Evidence), $\Delta$ln(BE),is also presented, relative to the best fitting model of the set. The top row of values for each case is assuming a homogeneous terminator cloud, with the lower row also retrieving a cloud fraction. The case with the highest Bayesian evidence is highlighted in green. The T18 model is highlighted in red due to the poor quality of the spectral fit.} 
	\label{table_189_results}
	\begin{tabular}{lcccc} 
		\hline
		Property & B17 & T18 & F18 & P19\\
		\hline
		Log(H$_2$O) & -4.94$^{+0.47}_{-0.25}$ & \textcolor{red}{-5.56$^{+0.23}_{-0.25}$} & -5.02$^{+0.21}_{-0.21}$ & -4.98$^{+0.25}_{-0.23}$ \\
		\\
		& -4.72$^{+0.32}_{-0.28}$ & \textcolor{red}{-7.50$^{+0.39}_{-0.32}$} & -4.94$^{+0.25}_{-0.22}$ & \textcolor{OliveGreen}{-3.58$^{+0.50}_{-0.49}$}\\
		\\
		Scat index & 6.34$^{+0.54}_{-2.39}$ & N/A & 6.37$^{+0.34}_{-0.34}$ & 6.47$^{+0.52}_{-0.38}$ \\
		\\
		& 8.13$^{+1.44}_{-1.12}$ & N/A & 6.62$^{+0.26}_{-0.35}$ & \textcolor{OliveGreen}{10.29$^{+1.41}_{-1.12}$} \\
		\\
		Top pressure & -6.56$^{+1.36}_{-0.99}$ & \textcolor{red}{0.28$^{+0.46}_{-0.47}$} & N/A & 0.34$^{+0.46}_{-0.52}$\\
		\\
		& -7.2$^{+0.98}_{-0.55}$ & \textcolor{red}{-3.45$^{+2.92}_{-2.99}$} & N/A & \textcolor{OliveGreen}{-2.29$^{+0.74}_{-0.62}$}\\
		\\
		Cloud fraction & N/A & N/A & N/A & N/A \\
		\\
		& 0.68$^{+0.17}_{-0.11}$ & \textcolor{red}{0.01$^{+0.01}_{-0.01}$} & 0.9$^{+0.07}_{-0.11}$ & \textcolor{OliveGreen}{0.61$^{+0.07}_{-0.07}$}\\
		\\
		$\Delta$ln(BE) & -6.8 &	\textcolor{red}{-41.9} &	-6.1 &	-6.3\\
& -6.4 &	\textcolor{red}{-188.9} &	-8 & \textcolor{OliveGreen}{0} \\
		\hline
	\end{tabular}
\end{table}

\begin{table}
	\centering
	\caption{Key retrieval results for HD 209458b, as Table~\ref{table_189_results}.} 
	\label{table_209_results}
	\begin{tabular}{lcccc} 
		\hline
		Property & B17 & T18 & F18 & P19\\
		\hline
		Log(H$_2$O) & -4.89$^{+0.23}_{-0.23}$& -5.02$^{+0.17}_{-0.15}$ & -5.11$^{+0.16}_{-0.15}$ & -4.95$^{+0.24}_{-0.19}$ \\
		\\
		& -5.15$^{+0.12}_{-0.11}$ & -5.19$^{+0.14}_{-0.14}$ & -5.18$^{+0.13}_{-0.12}$& \textcolor{OliveGreen}{-5.06$^{+0.17}_{-0.17}$}\\
		\\
		Scat index & 3.69$^{+5.16}_{-2.64}$ & N/A & 5.08$^{+1.31}_{-1.36}$ & 8.79$^{+3.49}_{-5.33}$ \\
		\\
		& 3.12$^{+2.48}_{-1.95}$& N/A & 4.84$^{+1.33}_{-1.17}$& \textcolor{OliveGreen}{8.23$^{+3.62}_{-4.07}$}\\
		\\
		Top pressure & -0.65$^{+0.26}_{-2.95}$ & -0.79$^{+0.37}_{-0.13}$ & N/A & -0.61$^{+0.16}_{-0.11}$\\
		\\
		& -2.89$^{+0.81}_{-1.05}$ & -3.64$^{+3.01}_{-2.88}$ & N/A & \textcolor{OliveGreen}{-5.6$^{+2.14}_{-1.6}$}\\
		\\
		Cloud fraction & N/A & N/A & N/A & N/A \\
		\\
		& 0.37$^{+0.05}_{-0.05}$ & 0.44$^{+0.07}_{-0.08}$ & 0.33$^{+0.06}_{-0.05}$ & \textcolor{OliveGreen}{0.39$^{+0.06}_{-0.08}$}\\
		\\
		$\Delta$ln(BE) &-10.9 &	-9.4 &	-7.3 &	-9\\
&-5.3&-1.3 &-4&\textcolor{OliveGreen}{0}\\
		\hline
	\end{tabular}
\end{table}

\subsection{HD 209458b}
\label{hd209_results}
Model fits to HD 209458b proved more straightforward than for HD 189733b due to the lack of a substantial scattering slope in the optical part of the spectrum. The retrieved temperature is in the range expected for the planet, even when a less restrictive temperature prior is used. The T18 model is also able to produce a good fit to the data in this case. 

\begin{figure*}
	\centering
	\includegraphics[width=\textwidth]{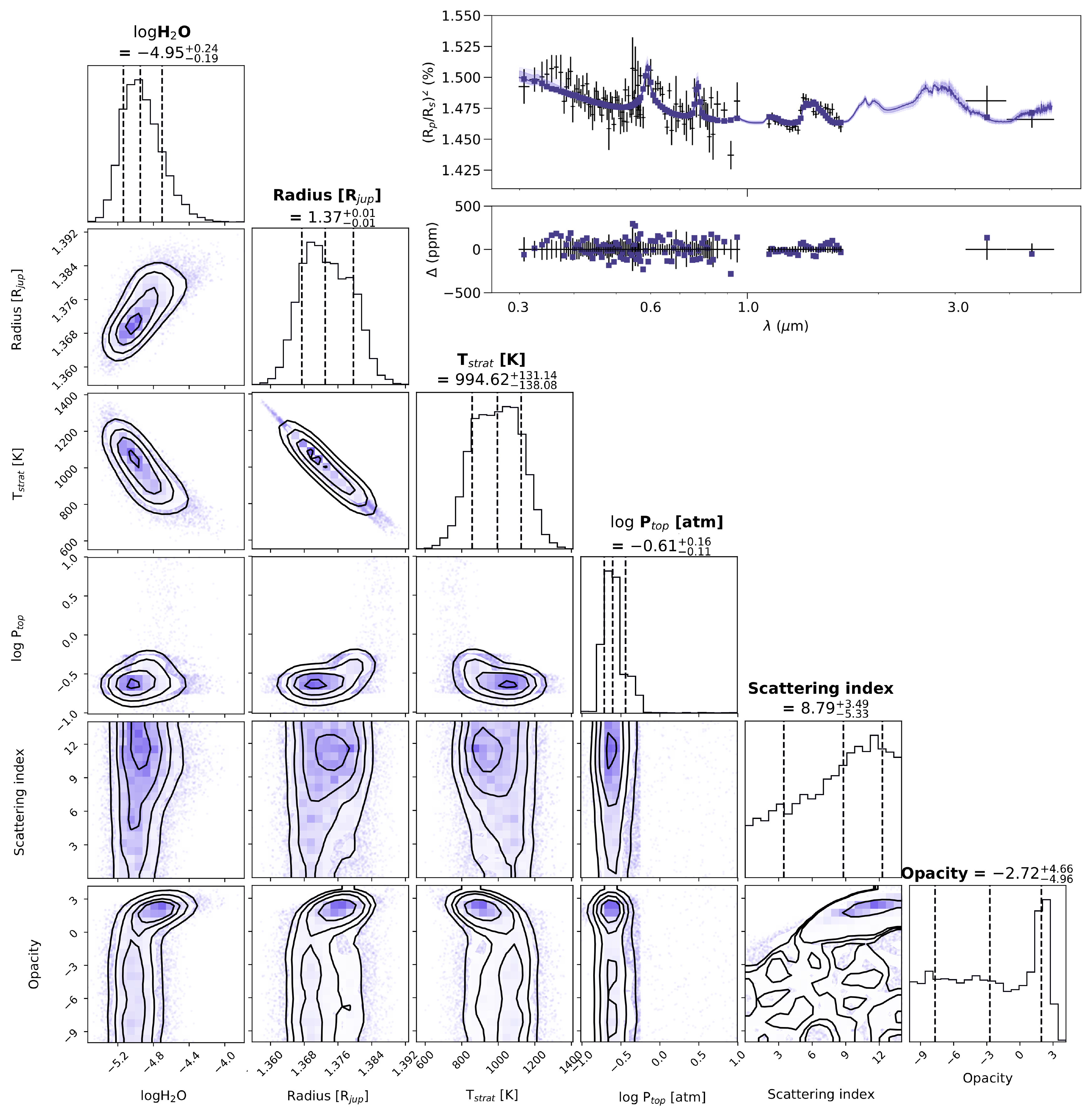}
    \caption{Posterior probability distributions (excluding alkali metals) and best-fit spectrum for HD 209458b using the P19 cloud model with 100$\%$ terminator cloud cover. The spectrum is generated using the median values of the posterior distributions. Shaded regions on the spectrum plot indicate the 1 (darker) and 2 (paler) $\sigma$ envelopes.}
    \label{hd209_p19}
\end{figure*}

\begin{figure*}
	\centering
	\includegraphics[width=\textwidth]{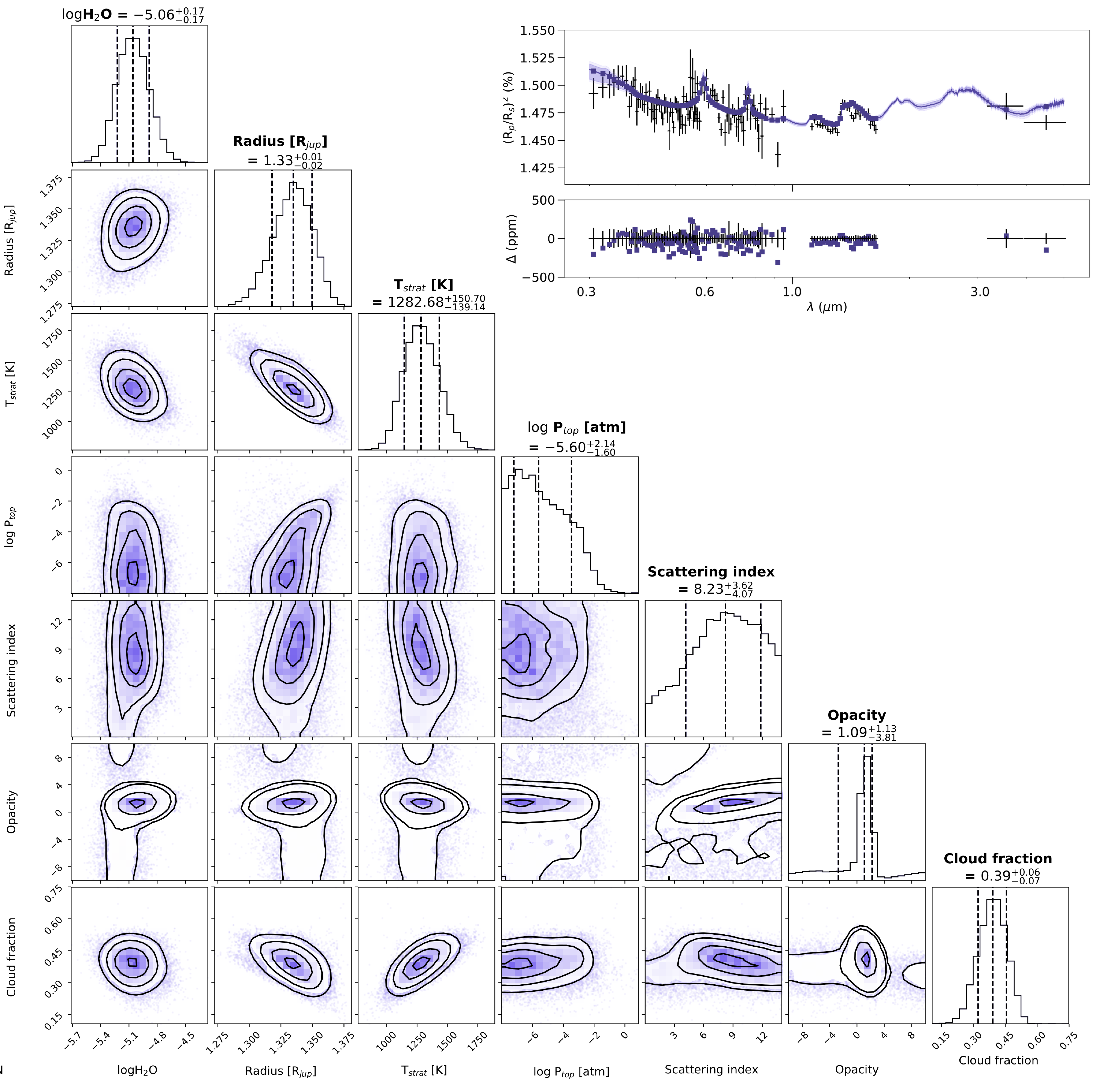}
    \caption{As Figure~\ref{hd209_p19}, but including a fractional cloud parameter.}
    \label{hd209_p19_cfrac}
\end{figure*}

\begin{figure*}
	\centering
	\includegraphics[width=\textwidth]{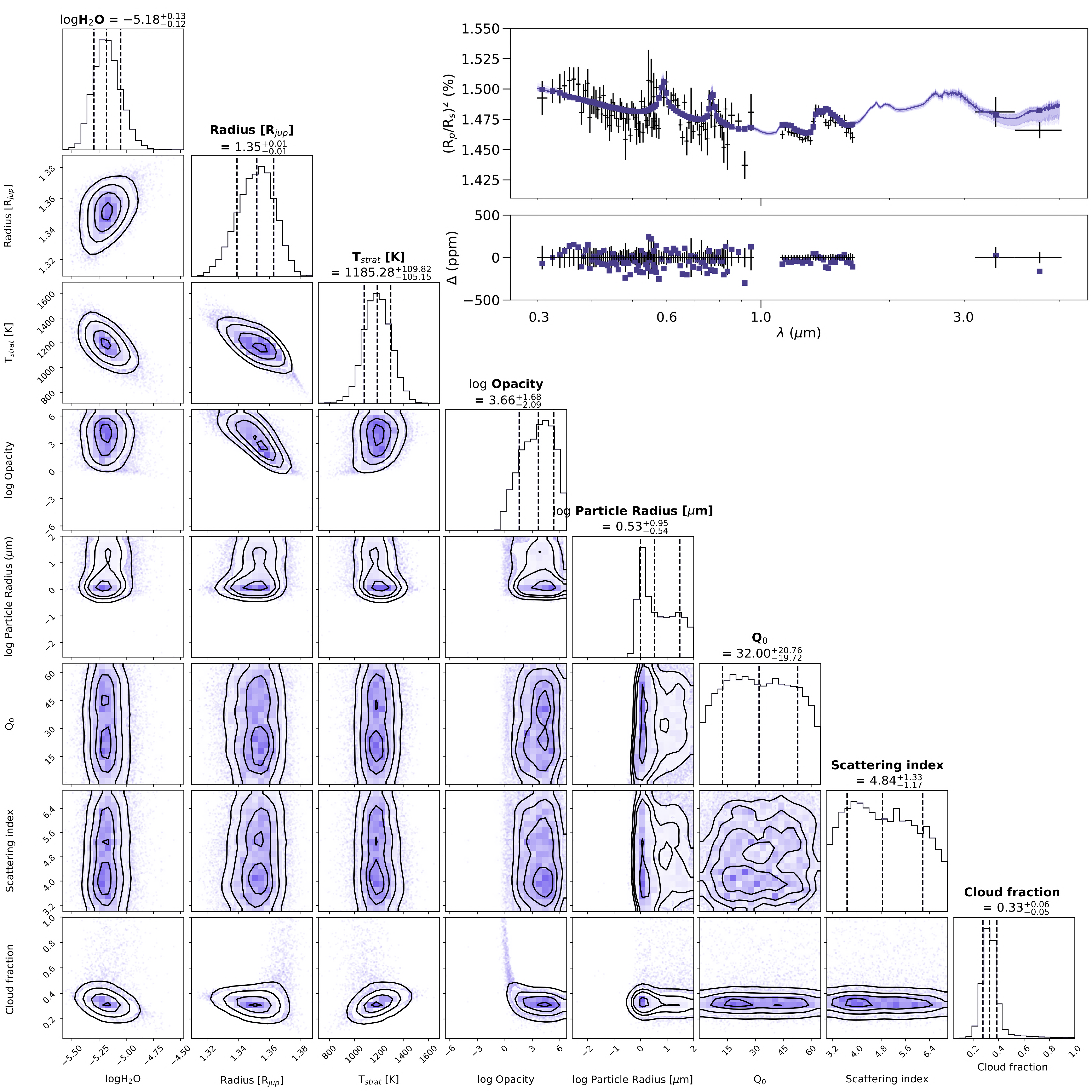}
    \caption{Posterior probability distributions (excluding alkali metals) and best-fit spectrum for HD 209458b using the F18 cloud model with fractional terminator cloud cover. The spectrum is generated using the median values of the posterior distributions. Shaded regions on the spectrum plot indicate the 1 (darker) and 2 (paler) $\sigma$ envelopes.}
    \label{hd209_f18_cfrac}
\end{figure*}

The key results for HD 209458b are presented in Table~\ref{hd209_results}. As for HD 189733b, the best fitting model is highlighted in green, and tests are run for both homogeneous and heterogeneous cloud. 

Again, the retrieved H$_2$O abundance is consistently sub solar across all models tested, at around 10 ppmv. Unlike the HD 189733b case, however, the evidence is more strongly in favour of a grey cloud deck, as both T18 and P19 models have top pressures that would make the grey cloud deck visible (see for example Figure~\ref{hd209_p19}). By contrast, the top pressure of the B17 model is higher than for HD 189733b, indicating that the upper part of the atmosphere is more likely to be cloud-free. 

All four models agree well on cloud fraction for the heterogeneous cloud case, and all four also show a substantial improvement in the goodness of fit when the cloud fraction parameter is included, with Bayesian evidence indicating that fractional cloud is strongly favoured. With fractional cloud included, the cloud top pressure for the T18 and P19 models is reduced (compare e.g. Figure~\ref{hd209_p19_cfrac} with Figure~\ref{hd209_p19}), as partial cloud high up has a similar observational signature to opaque cloud lower down. 

F18 and P19 have higher scattering indices than B17, but in the case of P19 there is relatively low opacity for the upper cloud/haze above the opaque grey cloud, as in this case the grey cloud dominates the signal. 

For the case where fractional cloud is included, the T18 and P19 models have similar evidence, whereas B17 and F18 are strongly and moderately disfavoured with respect to P19. T18 and P19 both include an opaque cloud deck, reinforcing the fact that HD 209458b is likely to have grey cloud.

\section{Discussion}
\label{discussion}
\begin{figure*}
	\centering
	\includegraphics[width=\textwidth]{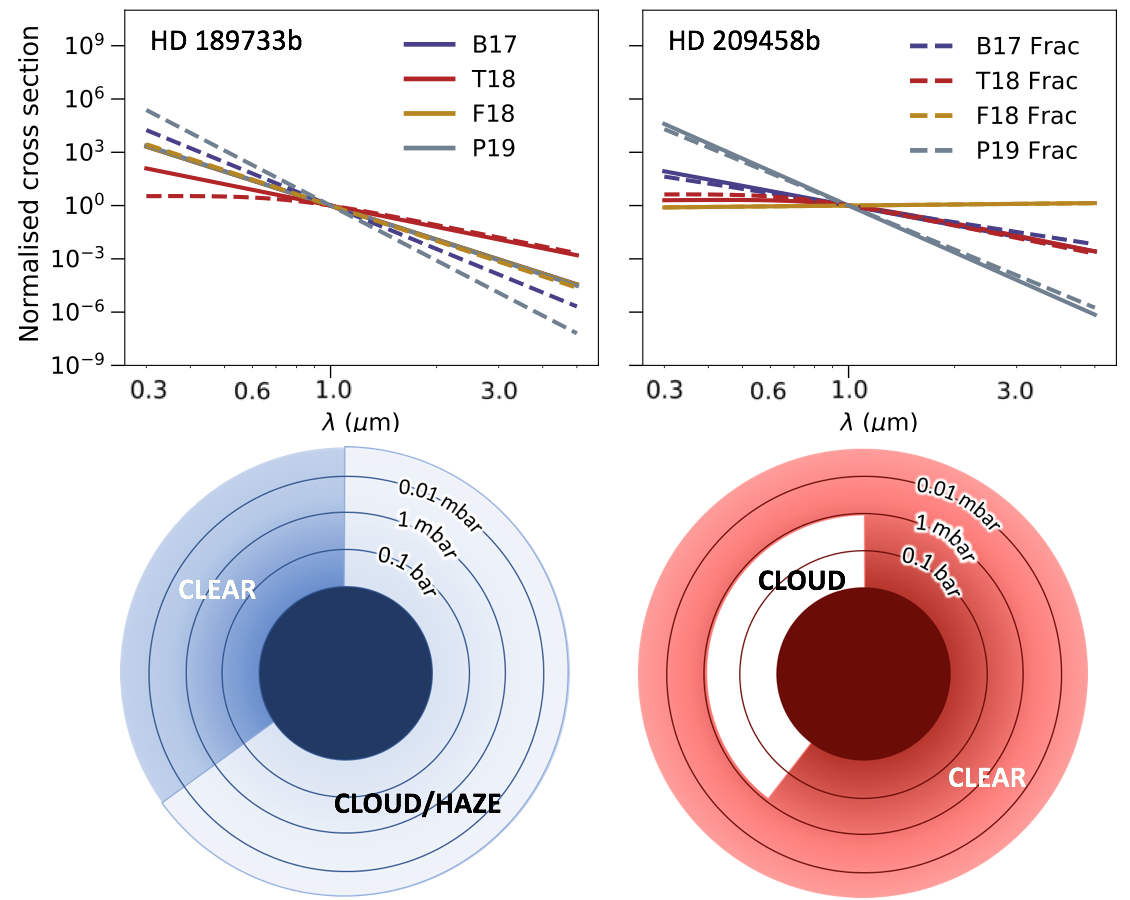}
    \caption{I show here a visual summary of the retrieved cloud structure and scattering behaviour. The cloud structure as shown emerges from the combined information of all four models, which indicates that HD 189733b has small-particle aerosols that cover at least 60$\%$ of the terminator, reaching to low pressures, but no grey cloud; and HD 209458b has opaque grey cloud deep in the atmosphere covering around 40$\%$ of the terminator. The top panels show the wavelength dependence of the cloud extinction cross section for each model case. The cross sections are normalised to the values at 1 $\upmu$m for comparison. The solid lines indicate the result for homogeneous cloud and the dotted lines for fractional cloud cover.}
    \label{cloud_summary}
\end{figure*}

Despite the fact that the four models include different ways of parameterising the cloud, for the most part a coherent picture emerges. These are visually summarised for each planet in Figure~\ref{cloud_summary}. 

For both cases, the P19 model emerges as the one that provides the best fit to the observed spectra. The combination of grey cloud and overlying haze with a tunable scattering index seems to provide the most flexibility.

Below, I discuss the major findings about H$_2$O abundance, cloud scattering properties and cloud location for each planet. I contrast the findings with those presented in the P19 paper. A similar comparison for the results from F18 and T18 is not relevant, as these two papers dealt with only a limited subset of the data considered here. 

\subsection{H$_2$O abundance}
\label{h2o}
For both planets, the retrieved H$_2$O abundance is generally very robust to different assumptions about cloud, echoing similar findings by \cite{mai19}. The exception to this is the heterogeneous cloud P19 model for HD 189733b, which has a retrieved log(H$_2$O) abundance of -3.58, meaning the H$_2$O volume mixing ratio is an order of magnitude higher than for all other models. This is also substantially higher than the retrieved value from the P19 paper itself (-5.04). The trade-off that causes this can be traced to differences in the retrieved cloud properties between the homogeneous model in this study, the results of P19 and the heterogeneous case here. The P19 study has a retrieved scattering index of $\sim8$, compared with 6.47 for the P19 parameterization in this work for the homogeneous case and 10.29 for the heterogeneous case. The higher the scattering index, the more rapidly the haze extinction efficiency drops off as a function of wavelength. To compensate for the lack of cloud opacity at longer wavelengths when the scattering index is high, the grey cloud top pressure for the heterogeneous P19 model in this work is reduced to -2.29 from 0.34 in the homogeneous model. The grey cloud moving higher up means that a higher H$_2$O abundance is required to fit the 1.4 micron feature in the WFC3 bandpass. 

Another key difference between this work and P19 is that this work also includes the bandpass integrated points from the \textit{HST}/NICMOS instrument, as published by \cite{pont13} after \cite{gibson12}. P19 did not include these points, and if they are removed and the fits using the P19 model are repeated, the solution is closer to the P19 published result, especially given the large error bars on the H$_2$O abundance (Table~\ref{P19_nonicmos}).\footnote{The fractional cloud case without NICMOS points for the P19 model has bimodal probability distributions for the radius and cloud top pressures (figures in online material). The result of this is that the spectrum generated using the median values actually doesn't provide a good fit, as the median falls in between the probability maxima. The maximum likelihood solution produces a spectrum that provides a much better fit to the observation.} 

\begin{table}
	\centering
	\caption{Comparison of retrieved results without the HD 189733b NICMOS data points from this work and P19. Error bars are not quoted for P19 values where they are not specified in detail in the paper. Values without error bars are as read from probability distribution histograms.} 
	\label{P19_nonicmos}
	\begin{tabular}{lccc} 
		\hline
		Property & 100\% cloud & Fractional cloud & P19 \\
		\hline
		Log(H$_2$O) & -5.21$^{+0.21}_{-0.21}$ & -4.23$^{+0.79}_{-0.72}$ & -5.04$^{+0.46}_{-0.30}$  \\
		\\
		Scat index & 6.49$^{+0.48}_{-0.37}$ & 10.52$^{+1.79}_{-1.84}$ & 7.75\\
		\\
		Top pressure & 0.45$^{+0.37}_{-0.38}$ & -1.36$^{+1.72}_{-1.22}$ & 0.4 \\
		\\
		Cloud fraction & N/A & 0.58$^{+0.11}_{-0.08}$ & 0.67 \\
		\hline
	\end{tabular}
\end{table}

All H$_2$O abundance values for HD 209458b are consistent within 1$\sigma$ for homogeneous and heterogeneous cloud models, and within 2$\sigma$ between homogeneous and heterogeneous models. The H$_2$O abundance is also consistently subsolar, as for HD 189733b, compared with the value predicted by \cite{moses11} of -3.45.  The retrieved H$_2$O abundance from the P19 paper is -4.66$^{+0.39}_{-0.30}$, slightly higher than the values presented here, but the probability distributions overlap at the 1$\sigma$ level. 

A likely explanation for any remaining differences between this work and the P19 paper is our different approaches to temperature profile parameterisation. In this work, I use a simple model assuming an isotherm plus an adiabat, whereas the more complex 6-parameter model of P19 allows greater freedom in the structure of the T-p profile. Whilst in the region of greatest sensitivity the T-p profiles are consistent with each other, variation in the lower atmosphere (where there is little constraint) could alter the deep atmosphere scale height, which could in turn affect the other retrieved properties.

\subsection{Scattering properties}
\label{scattering}
\cite{pinhas17} investigate the most representative scattering slope index for a range of possible cloud compositions with different particle sizes, and find that values of around 6 can be achieved with modal sizes of 0.01 $\upmu$m for Na$_2$S and ZnS, and 0.1 $\upmu$m for MnS. These slopes are only relevant across relatively narrow spectral ranges in the optical, and a full study of the effective cross sections for these species in \cite{pinhas17} shows that the curve somewhat flattens out in the near infrared. In reality, the variation of extinction cross section as a function of wavelength is much more complex than a simple power law relationship. 

However, it is clear that an extremely steep scattering slope is required to fit the optical spectrum for HD 189733b, that extends throughout the optical region. This slope does not appear to be achievable within any single species tested by \cite{pinhas17}. A secondary solution, indicated by the homogeneous B17 model (see Figure~\ref{hd189_b}), is that a steep slope is created by a detached haze layer high in the atmosphere that becomes optically thin at longer wavelengths. Other, more complex solutions could include layered clouds of different species. 

Previous explanations for the steep slope have also included unocculted starspots \citep{mccullough14}, although the data used in this work were in fact already corrected for unocculted starspots according to the method outlined by \cite{pont13}. However, this method was only able to account for the variable level of starspots, whilst the baseline spot coverage remains unknown and could still have an effect \citep{rackham18}. 

Our best hope for further understanding the cloud properties of HD 189733b is that JWST will uncover spectral signatures of a specific condensate in the infrared (e.g. \citealt{wakeford15}); at this point, it is unclear which effect of many is responsible for the steep optical slope. 

No such explanations need to be invoked for HD 209458b, as relatively flat spectra like this can be produced by any condensate with a large spread of particle sizes. Whilst retrieved scattering indices are high for the P19 model, this refers only to the upper layer of cloud which has a low optical depth - it is the grey cloud deck for this and for the T18 model that provides the majority of the cloud opacity. The scattering index for B17 and F18 is much lower for HD 209458b than for HD 189733b. I therefore conclude that the cloud on HD 209458b is consistent with cloud containing large aerosol particles. 

\subsection{Location of cloud}
\label{location}
Again, the picture is more complicated for HD 189733b than for HD 209458b. Whilst all cases are consistent with a low top pressure for a small particle haze (from the B17 model), and a relatively high top pressure for any grey cloud that might be present (from T18 and P19), there is ambiguity about the spatial location of the cloud. Whilst with the best fitting P19 model there is a clear improvement when a fractional cloud parameter is included, this is not the case for the B17 or F18 models. 

The retrieved cloud fraction from the P19 model is consistent with the value from the P19 paper itself, although the probability distribution from the P19 paper is double-peaked and also has a secondary maximum close to 1.0. This reflects the spread of cloud fraction values from the other models tested in this work. There is also substantial degeneracy between the grey cloud top pressure and cloud fraction for the P19 model, as when a fractional cloud parameter is included the top pressure decreases from greater than 1 bar\footnote{NEMESIS uses units of atmospheres for pressure. 1 atm  = 1.01325 bar.} to less than 10 mbar (Table~\ref{table_189_results}). The lower value here is not consistent with the result from P19, that has the grey cloud top pressure at around 1 bar, but removing the NICMOS points reduces the retrieved pressure in this work to a somewhat closer value (Table~\ref{P19_nonicmos}). In any event, the degeneracy first pointed out in \cite{line16} between global and patchy cloud is evident here. 

On the other hand, there is strong and consistent evidence that HD 209458b has a terminator with only $\sim$40$\%$ cloud coverage. This is also in reasonable agreement with the value from the P19 paper, which has a cloud fraction of 51$\%$, with the probability distribution overlapping with the one in this work. 

The location of the cloud top for HD 209458b is a little more ambiguous than for HD 189733b, because once again there is substantial degeneracy between the cloud fraction and the cloud top pressure. Lower cloud top pressures are permitted for fractional terminator cloud coverage, as their relative effects on the cloud opacity cancel out. Indeed, for the P19 model the top of the grey cloud deck could be very high in the atmosphere, which may be somewhat implausible as large particles would be unlikely to be lofted to pressures of order a few $\upmu$bar. However, it should be noted that the constraint for the P19 model is only an upper limit on the cloud top pressure. 

HD 189733b and HD 209458b have fairly similar equilibrium temperatures; HD 209458b is around 300 K warmer. They key difference relevant here is probably that HD 209458b is a highly inflated planet, whereas HD 189733b has a relatively high gravity for a hot Jupiter. This might explain why the cloud on HD 189733b appears to be composed of much smaller particles than on HD 209458b, as large particles would be more likely to rain out in a higher gravity environment. 

Assuming that the cloud on HD 189733b and HD 209458b may be made of the same substance, the presence of cloud at only part of the terminator on HD 209458b may be due to its higher equilibrium temperature. Whilst the temperature may be suitable for cloud on both terminators for HD 189733b, the dayside and evening terminator of HD 209458b may be too warm for cloud to persist. 

\subsection{Alkali metals}

In this paper, I have focused on the retrievals of H$_2$O abundance and cloud properties rather than the alkali metals. Na is detected in both atmospheres, whilst only an upper limit for K can be obtained for HD 189733b. In general, the retrieved abundances are consistent between cloud models, although the presence of fractional cloud affects the Na abundance for HD 189733b.

For HD 189733b (without fractional cloud), Na and K abundances are approximately 600 and 0.03 ppmv respectively; when fractional cloud is included, the Na abundance is reduced to $\sim$ 10 ppm. For HD 209458b, Na and K abundances are approximately 200 and 4 ppmv. Whilst the K abundances are reasonably aligned with expectations (solar abundance from \cite{asplund09} is $\sim$0.1 ppmv), the Na abundances are surprisingly high (several hundred ppmv, as opposed to $\sim$2 ppmv from \cite{asplund09}). The full Na and K retrieved results are included in the online repository that accompanies this paper.

As mentioned in \cite{barstow17}, the limited resolution of the k tables in the model present a challenge when fitting the centre of the alkali metal absorption bands, since these are observed using narrower wavelength bins than the rest of the spectrum. This is likely to be a factor in the somewhat unrealistic Na abundances from the retrieval, and means that the retrieved abundances should not be relied on. This is unlikely to affect other retrieved properties, since the continuum in both spectra is dominated by cloud rather than by the alkali line wings.

\section{Conclusions}
This paper has explored a range of cloud parameterisations that exist in the literature and applied them to spectra of HD 189733b and HD 209458b. I find that, whilst each model has different approaches to representing cloud, the retrieval results taken together present a surprisingly coherent and holistic picture of each planet. HD 189733b most likely has cloud made of small particles spread throughout the region of the atmosphere to which we are sensitive, whereas HD 209458b displays a thicker cloud with larger particles that is restricted both to lower regions of the atmosphere and to roughly 40$\%$ of the terminator. Both planets have H$_2$O abundances that are consistently retrieved to be subsolar, confirming the previous findings of \cite{barstow17} and \cite{pinhas19}. 

Despite their relatively similar equilibrium temperatures, and apparently similar chemistry extrapolated from their H$_2$O abundances, the cloud properties of the two planets indicate very different regimes. The lack of large particles on HD 189733b may be attributable to its higher gravity, whereas in the partially cloudy terminator on HD 209458b we may be seeing the effect of a slightly warmer eastern (evening) terminator. 

We now have access to an ever increasing number of hot Jupiters with spectra covering the optical and infrared, so this type of analysis can and should be expanded to cover a broader range of targets. The James Webb Space Telescope will provide more insight by probing further into the infrared, and could potentially reveal what these mysterious clouds are made of.

\section*{Acknowledgements}

I thank the anonymous referee for a very thorough report and several helpful suggestions that substantially improved the clarity of the paper. I acknowledge the support of a Royal Astronomical Society Research Fellowship. I thank Pat Irwin for the use of NEMESIS and Dan Foreman-Mackey for the use of the corner.py routine, which is available to download on GitHub: https://github.com/dfm/corner.py. I am grateful to the Center for Open Science for their provision of the free Open Science Framework data hosting service. 

\section*{Data Availability}
The data for HD 189733b and HD 209458b used in this paper are as published by \cite{sing16}. The spectra are available to download here: 

\url{https://pages.jh.edu/~dsing3/spectra/HD189733b_Sing_2015_Nature.csv}

\url{https://pages.jh.edu/~dsing3/spectra/HD209458b_Sing_2015_Nature.csv}



\bibliographystyle{mnras}
\bibliography{bibliography} 


\bsp	
\label{lastpage}
\end{document}